\documentclass[twocolumn,english,aps,prb,twocolum,superscriptaddress,natbib,bibnotes,amsmath,amssymb,floatfix,groupedaddress,footinbib]{revtex4-2}

\usepackage[colorlinks=true,citecolor=blue,linkcolor=magenta]{hyperref}

\usepackage[markup=nocolor, authormarkupposition=left]{changes} 
\usepackage{soul}
\usepackage[utf8]{inputenc}
\usepackage[english]{babel}
\usepackage{amsmath,amsfonts,amssymb}
\usepackage[T1]{fontenc}
\usepackage{url}

\usepackage{amsmath}
\usepackage{siunitx}
\usepackage[version=4]{mhchem}
\usepackage{amsfonts}
\usepackage{amssymb}

\usepackage{epstopdf}
\usepackage{graphicx}
\graphicspath{{./Figures/}}

\usepackage{changes}
\DeclareUnicodeCharacter{2212}{-}
\begin{document}

\title{Ultra-compact and efficient integrated multichannel mode multiplexer in silicon for few-mode fibers}

\author{Wu Zhou$^{1}$, 
        Zunyue Zhang$^{2,3}$, 
        Hao Chen$^{1}$, 
        Hon Ki Tsang$^{2}$, 
        and Yeyu Tong$^{1,\dagger}$}
\affiliation{
$^1$Microelectronic Thrust, Function Hub, The Hong Kong University of Science and Technology (Guangzhou), Guangdong, PR China\\
$^2$Department of Electronic Engineering, The Chinese University of Hong Kong, Shatin, New Territories, Hong Kong, PR China\\
$^3$School of Precision Instrument and Opto-Electronics Engineering, Tianjin University, Tianjin, PR China\\
}

\maketitle

\noindent\textbf{\noindent Space-division multiplexing (SDM) is one of the key enabling technologies to increase the capacity of fiber communication systems. However, implementing SDM-based systems using multimode fiber has been challenging with the need for compact, low-cost, and scalable mode de/multiplexer (DE/MUX). Here we present a novel integrated mode MUX for few-mode fibers (FMFs) which can launch up to eight spatial and polarization channels. The new design is composed of a two-dimensional multimode grating coupler (MMGC), highly compact mode size converters (MSCs), and adiabatic directional couplers (ADCs). Eight data lanes in FMFs can be selectively launched with integrated optical phase shifters. Experimental results reveal efficient chip-to-fiber coupling with peak efficiencies of -3.8 dB, -5.5 dB, -3.6 dB, and -4.1 dB for LP\textsubscript{01}, LP\textsubscript{11a}, LP\textsubscript{11b}, and LP\textsubscript{21b} modes, respectively. Meanwhile, the proposed design can efficiently couple all the degenerate LP modes in a two-mode FMF, allowing signal descrambling in the demultiplexer. Thanks to the use of integrated subwavelength Mikaelian lens for mode-independent field size conversion with loss $\leq\,\text{0.25 dB}$
, the total footprint of the MMGC and MSCs is only 35×35 $\mu$m\textsuperscript{2}. The proposed design shows great potential for densely integrated photonic circuits in future SDM applications.}

\section*{Introduction} 

\noindent{The} rapid expansion of new applications including cloud computing, virtual reality, and generative artificial intelligence technologies such as ChatGPT, has maintained the trajectory of exponential growth in datacenter traffic, and present new challenges for data center communication networks to meet the market demand. Space-division multiplexing (SDM)  has been widely investigated for optical fiber communications \cite{richardson_space-division_2013,winzer_making_2014,puttnam_space-division_2021}. By utilizing the independent data channels encoded on different orthogonal modes in a multimode fiber (MMF) or few-mode fiber (FMF), the communication capacity can be dramatically multiplied, which is referred to mode-division multiplexing (MDM). Significant progress has thus been made in recent years particularly in the specially designed SDM fibers \cite{kobayashi_1-pbs_2017,gregg_enhanced_2019}, fiber-based photonic lanterns \cite{velazquez-benitez_six_2015}, laser inscribed 3-dimensional waveguides \cite{gross_three-dimensional_2014}, multi-plane light conversion (MPLC) techniques \cite{labroille_efficient_2014,fontaine_laguerre-gaussian_2019}, and implementation of high-capacity communication systems \cite{soma_1016-peta-bs_2018,rademacher_peta-bit-per-second_2021,rademacher_1066_2020,liu_1-pbps_2022} with multiple-input multiple-output (MIMO) digital signal processing (DSP). 

To enable combination or separation of independent data streams in a FMF, mode multiplexers (MUXs) or demultiplexer (DEMUX) are essential to convert between single-mode signals and different mode patterns in the FMF. One of the key challenges in MDM-based communication systems is the monolithic integration of the mode DE/MUX interfaced with FMF, as opposed to the bulk optics technology used in MPLC demonstrations \cite{labroille_efficient_2014,fontaine_laguerre-gaussian_2019}. Photonic integration is essential for the deployment in low-cost and high-volume production and have potential for co-integration with photonic processors and high-speed optoelectronic transceivers \cite{shi_scaling_2020,bogaerts_programmable_2020,miller_device_2009,zhao_96-channel_2022,dai_silicon-based_2014,tong_experimental_2019,yang_multi-dimensional_2022}. However, it is challenging to establish efficient and reliable multichannel optical I/O between the 3D FMF and the planar lithographically fabricated photonic integrated circuits \cite{yang_multi-dimensional_2022,koonen_silicon_2012,ding_silicon_2013,wohlfeil_two-dimensional_2016,lai_compact_2018,tong_efficient_2019,watanabe_coherent_2020,shen_silicon-integrated_2020,zhang_low-crosstalk_2023,zhou_high_2022,zhang_-chip_2017}.

Thanks to the high-refractive-index contrast, silicon photonics can provide compact solutions for light field manipulation \cite{cheben_subwavelength_2018,luquegonzalez_ultracompact_2019,liu_arbitrarily_2019,zhang_nonparaxial_2020,xiang_metamaterial-enabled_2022}. Previously, 6 linearly polarized (LP) modes including LP\textsubscript{01}, LP\textsubscript{11a}, and LP\textsubscript{11b} modes in two orthogonal polarizations can be excited in a FMF by employing a novel diffraction grating coupler array on the silicon-on-insulator (SOI) platform. The reported coupling efficiency was less than -20 dB \cite{koonen_silicon_2012,ding_silicon_2013}. By optimizing the mode field matching conditions \cite{zhang_low-crosstalk_2023} or using a shifted polysilicon overlay \cite{zhou_high_2022}, the coupling efficiency can be improved to -1.36 dB and -2.21 dB for the LP\textsubscript{01} and LP\textsubscript{11} modes respectively \cite{zhou_high_2022}. But the number of spatial channels is still limited. We proposed a novel design approach and proved that the diffraction efficiency of two-dimensional (2D) multimode gratings can be similar to that of single-mode gratings in \cite{tong_efficient_2019}. The reported design shows an experimental coupling efficiency of -4.9 dB and -6.1 dB for LP\textsubscript{01} and LP\textsubscript{11} modes respectively in the two orthogonal polarizations, showing a good coupling efficiency while supporting four spatial channels. Using a blazing grating coupler array, it is possible to demultiplex six LP modes with a low coupling loss of -5.2 dB and -9.0 dB for the LP\textsubscript{01} and LP\textsubscript{11}, respectively \cite{watanabe_coherent_2020}. Equalizing mode-dependent loss in a DEMUX is difficult due to the unpredictable degenerate LP mode evolution and polarization rotation at the end of the FMF \cite{watanabe_coherent_2020,kreysing_dynamic_2014}. To overcome this issue and enhance the matching between high-order fiber modes and planar waveguide modes, a rectangular-core MMF was utilized in a MDM communication system \cite{yang_multi-dimensional_2022}. The inverse-designed coupler is feasible to support four spatial channels with a mode-dependent loss difference of less than 2.5 dB.

%%%%%%%%%%%%%%%%%%%%%%%%%%%%%%%%%%%%%%%%%%%%
% FIGURE 1 - principle graph
\begin{figure*}[!htp]
  \centering{
  \includegraphics[width = 0.97\linewidth]{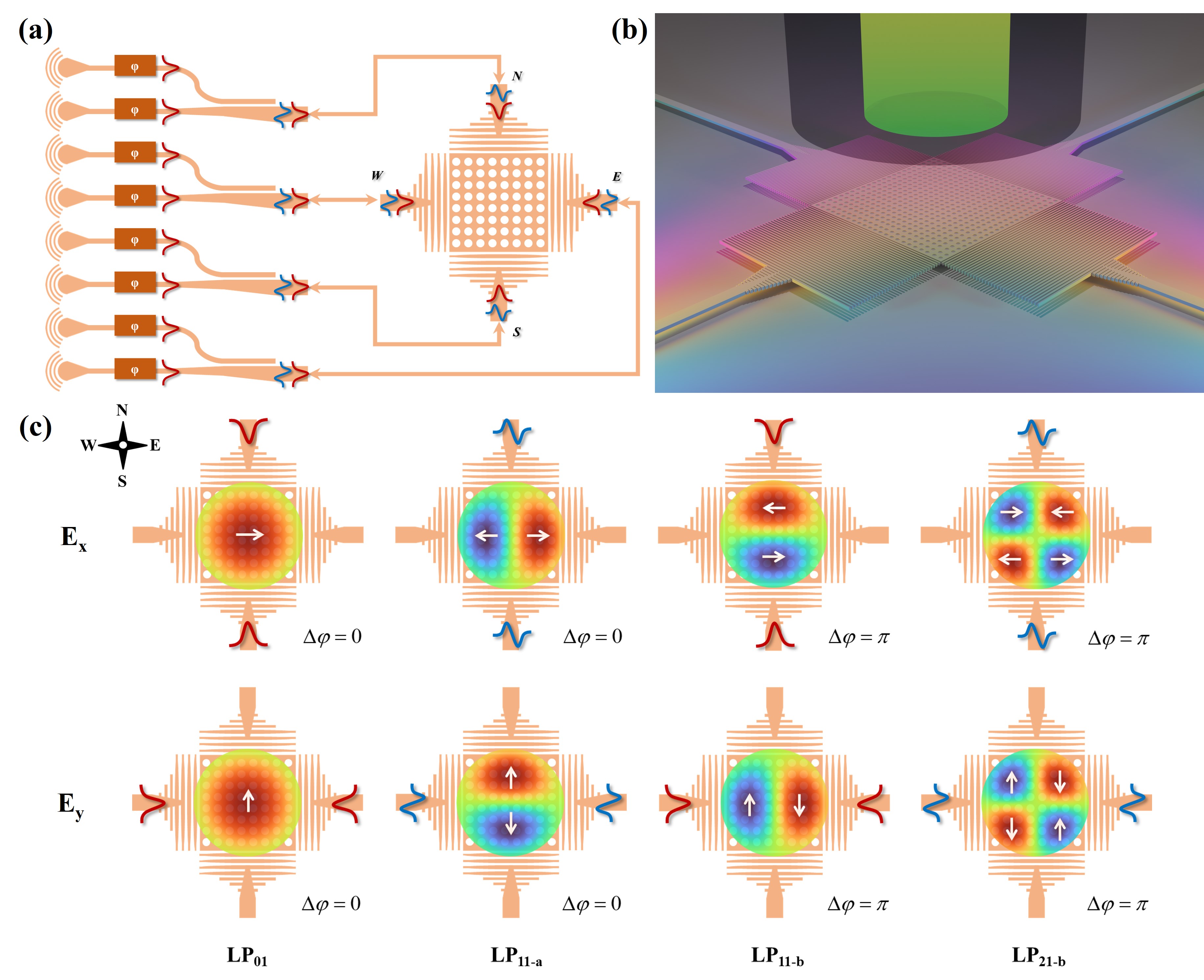}
  } 
    \caption{(a) Schematic illustrations of the integrated multimode DE/MUX, which consists of a 2D MMGC, four MSCs, and four ADCs. Eight heater-based optical phase shifters are needed for mode MUX. (b) A bird view of the MSCs and MMGC designed for perfectly vertical chip-to-fiber coupling. (c) Schematic illustration of selective mode launching for mode LP\textsubscript{01}, LP\textsubscript{11a}, LP\textsubscript{11b}, LP\textsubscript{21b} in a FMF via proposed design.
  }
 \label{Figure1}
\end{figure*} 
%%%%%%%%%%%%%%%%%%%%%%%%%%%%%%%%%%%%%%%%%%%%

While photonic integration approach has significant potential benefits, a reliable, efficient, and integrated multichannel mode DE/MUX solution is still lacking. To tackle this problem, in this work, we present a novel integrated multichannel mode MUX for FMFs. The proposed design can launch up to 8 spatial channels in a FMF including all the degenerate modes in LP\textsubscript{01}, LP\textsubscript{11}, and two degenerate modes in the LP\textsubscript{21} mode group. In the experiment, the integrated mode MUX shows a minimum chip-to-fiber coupling loss of -3.8 dB, -5.5 dB, -3.6 dB, -4.1 dB for LP\textsubscript{01}, LP\textsubscript{11a}, LP\textsubscript{11b}, and LP\textsubscript{21b} modes, respectively. Selective excitation is achieved for the four LP modes in the two orthogonal polarizations with integrated phase shifters. The reported design also maintains good fiber-to-chip efficiency for all degenerate modes in a 2-mode FMF, allowing mode descrambling such as MIMO processing for the DEMUX. Moreover, the 2D multimode grating coupler (MMGC) and mode size converters (MSCs) fabricated in this study are extremely compact showing a total footprint of only 35×35 $\mu$m\textsuperscript{2}, which is mainly attributed to the mode-independent field size conversion enabled by the subwavelength Mikaelian lens. The proposed design demonstrates the potential of photonic integrated solutions for many SDM applications, such as data transmissions \cite{richardson_space-division_2013,winzer_making_2014,puttnam_space-division_2021}, quantum information processing \cite{xavier_quantum_2020,carolan_universal_2015}, imaging \cite{cizmar_exploiting_2012}, and spectroscopy \cite{wan_high-resolution_2015}.

\section*{Design and Simulation Results}
The proposed design works for two-mode FMF and four-mode FMF provided by OFS. The two-mode FMF is capable of supporting LP\textsubscript{01} and LP\textsubscript{11} modes, while the four-mode FMF is designed to accommodate LP\textsubscript{01}, LP\textsubscript{02}, LP\textsubscript{11}, and LP\textsubscript{21} modes (see Figure S1). A graded-index profile results in small absolute differential mode group delay, which is $\leq 0.2\,\text{ps/m}$ in the two-mode FMF, and $\leq 0.4\,\text{ps/m}$ in the four-mode FMF.

The proposed integrated mode DE/MUX for FMF is composed of a 2D multimode grating coupler (MMGC), four mode-size converters (MSCs), and four tapered adiabatic directional couplers (ADCs) as depicted in Figure \ref{Figure1}a. Tapered ADCs are utilized for on-chip mode de/multiplexing of the TE\textsubscript{0} mode and TE\textsubscript{1} mode \cite{ding_-chip_2013,dai_silicon_2013} (see Supporting Information S1). Additionally, heater-based optical phase shifters are also necessary for selective mode excitation as a mode MUX. Intrinsic to its design, the 2D MMGC shown in Figure \ref{Figure1}b serves as both a polarization combiner in the mode MUX and a polarization splitter in the mode DEMUX.

The selective mode launching of LP modes is depicted in Figure \ref{Figure1}c, which can be explained as follows: The mode field profile of the LP\textsubscript{01} mode in a FMF has only one maximum intensity point. A 2D grating coupler can be utilized to diffract the fundamental quasi transverse-electric mode (TE\textsubscript{0}) in the four orthogonally placed waveguides. By using the two counterpropagating TE\textsubscript{0} modes in the same polarization with no relative phase difference, LP\textsubscript{01} modes can be selectively excited. For the higher-order fiber modes, e.g. LP\textsubscript{11} and LP\textsubscript{21} modes, additional relative phase shift is required to match the diffracted optical field with the field distribution in FMF. As illustrated in Figure \ref{Figure1}c, LP\textsubscript{11a-x} mode can be excited by the TE\textsubscript{1} modes from south and north with no relative phase delay allowing the spots to be merged; LP\textsubscript{11b-x} can be obtained by the TE\textsubscript{0} modes from south and north with a relative phase shift of $\pi$. Similarly, LP\textsubscript{11a-y} and LP\textsubscript{11b-y} can be selectively launched using the TE\textsubscript{0} and TE\textsubscript{1} modes from west and east with appropriate phase shifts. For the LP\textsubscript{21} mode group, LP\textsubscript{21b-x} mode can be excited by the TE\textsubscript{1} modes from south and north in the same phase state. TE\textsubscript{1} modes from the west and east can be used for launching the LP\textsubscript{21b-y} mode. A total of eight spatial modes can thus be selectively coupled via the proposed 2D MMGC.

The same design configuration can also be used as a mode DEMUX in the future. Selective decoupling only occurs when the fiber LP modes are of high modal purity and polarizations are precisely aligned. However, in practical scenarios, mode evolution of the LP modes and polarization rotation cannot be avoided in a circular-core FMF. As a result, unpredictable field patterns are generated at the fiber-to-chip coupling end \cite{watanabe_coherent_2020,kreysing_dynamic_2014}. Nevertheless, for a two-mode FMF, our proposed design can receive all the degenerate modes in LP\textsubscript{01} and LP\textsubscript{11} group and convert them into 8 single-mode spatial channels on chip with preserved optical energy. While polarization crosstalk and modal crosstalk can result in signal degradation, it is possible to mitigate this issue through digital MIMO signal processing \cite{ryf_mode-division_2012} or the use of Mach-Zehnder interferometer (MZI) meshes \cite{annoni_unscrambling_2017}, as demonstrated in previous studies.

\subsection{Multimode Grating Coupler Design}
\begin{figure*}[!!htp]
  \centering{
  \includegraphics[width = 0.97\linewidth]{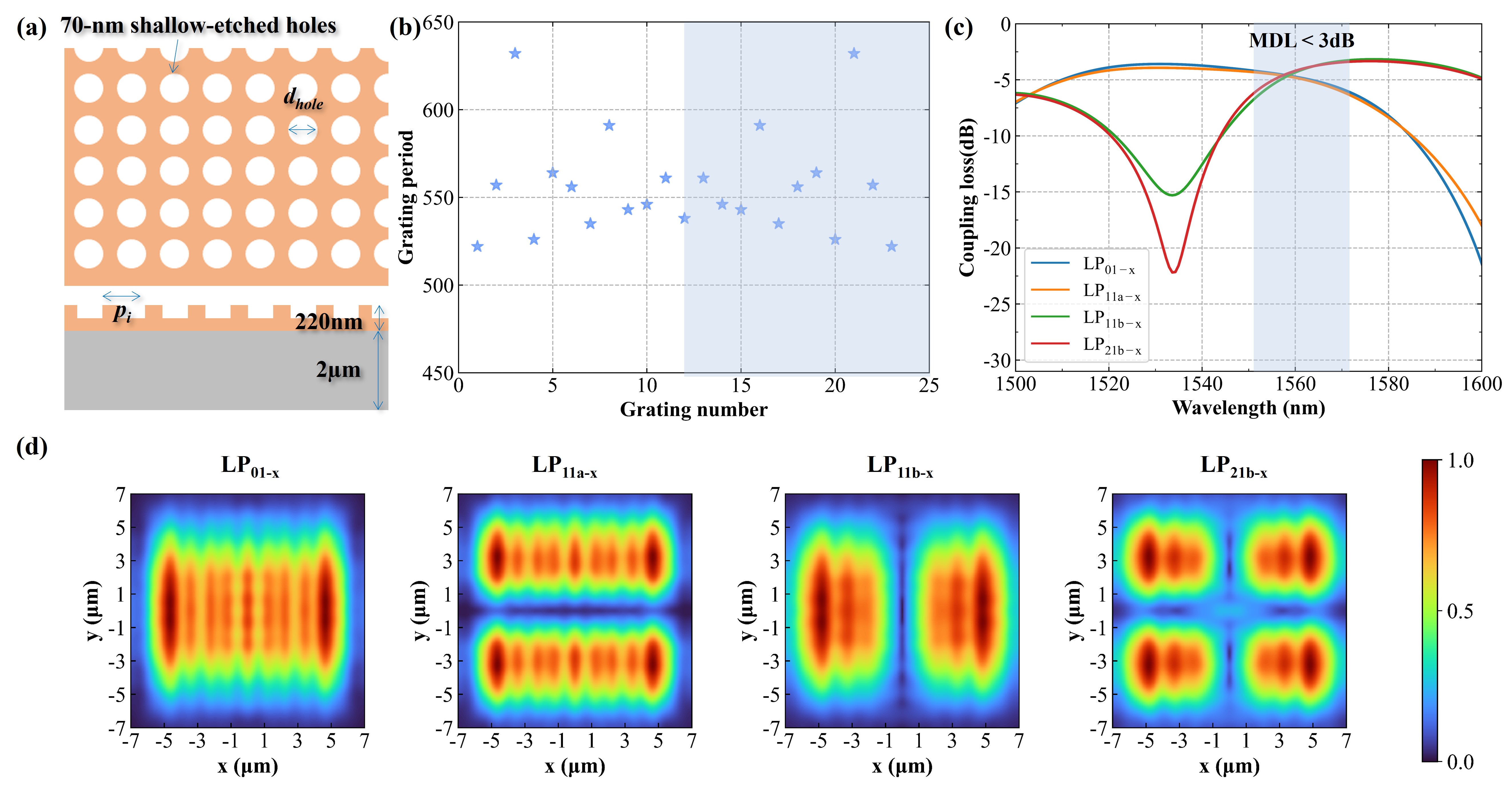}
  } 
    \caption{(a) Cross section and structural parameters of the MMGC. (b) Period of the chirped subwavelength gratings. (c) Simulated coupling loss spectra for different spatial modes by 3D-FDTD simulation. (d) Diffracted optical field intensity profile obtained in 3D-FDTD simulation.
  }
 \label{Figure2}
\end{figure*}

The proposed multimode grating is efficient for both TE\textsubscript{0} and TE\textsubscript{1} modes due to their comparable effective indices in the wide slab waveguide, which has a width exceeding 10 $\mu$m. Our design utilizes a completely vertical coupling approach to decrease spatial dependent loss and eliminate angled fiber polishing. To make sure that the performance is consistent for all modes coming from the four slab waveguides in various directions, the design utilizes a symmetrical structure. As presented in Figure \ref{Figure2}a, the SOI wafer comprises a top silicon layer with a thickness of 220 nm and a buried-oxide layer with a thickness of 2 $\mu$m. A shallow etching process with a depth of 70 nm is used to create subwavelength holes as the low-index region in the grating, and a full etching process is needed later to make the strip waveguides.

It is known that grating couplers usually have a small coupling angle with respect to the chip surface normal to avoid second-order Bragg reflection into the waveguides. As a result, advanced designs such as chirped gratings \cite{chen_fabrication-tolerant_2008,tong_efficient_2018}, are necessary for achieving efficient and perfectly vertical coupling. However, optimizing the chirping conditions for a 2D grating coupler is challenging. The 3D finite element simulations must consider the large fiber core and cladding as well as the sub-wavelength holes, making them computationally intensive. Estimating the figure-of-merit (FOM) of different fiber modes requires significant computing resources, which makes it difficult for optimization algorithms to converge.

To solve this problem, effective medium theory (EMT) was used to simplify the modelling into a 2D finite element simulation. While 2D simulations cannot accurately evaluate the coupling efficiencies for all fiber modes in the two orthogonal polarizations, they can easily obtain the coupling performance of the fundamental TE mode, which is sufficient for evaluating the FOM during the optimization process. Essentially, the MMGC is wide and symmetric for the two orthogonal polarizations, thus all the TE\textsubscript{0} and TE\textsubscript{1} modes in the orthogonal polarizations should exhibit similar or equivalent effective indices. Assuming good mode size matching, the coupling performance of the higher order fiber modes should be directly proportional to the results of the fundamental mode in 2D simulations. As a result of this, the computing resources needed to evaluate the FOM can be greatly reduced, making it possible for more sophisticated optimization algorithms, such as genetic algorithm in this work, to search the best design parameters. The second order EMT can be found in Supporting Information S2.

Figure \ref{Figure2}b presents the length of each grating period with a constant etched hole diameter d\textsubscript{hole} of 343 nm. Performance of the chirped MMGC is verified using 3D finite-difference time-domain (FDTD) simulation. By configuring the appropriate phase shift as depicted in Figure \ref{Figure1}c, the proposed MMGC can selectively excite various fiber LP modes. Figure \ref{Figure2}c shows the coupling loss spectra for different x-polarized LP modes. The symmetrical design guarantees equivalent performance for the LP mode in the y polarization. LP\textsubscript{01} and LP\textsubscript{11a} mode have a minimum coupling loss of -3.61 dB and -3.94 dB at a center wavelength of 1532 nm, with a 1-dB spectral bandwidth of about 47 nm. The minimum coupling loss for LP\textsubscript{11b} mode and LP\textsubscript{21b} mode are -3.17 dB and -3.75 dB, at 1577 nm respectively. The 1-dB spectral bandwidth for LP\textsubscript{11b} mode and LP\textsubscript{21b} mode are around 35 nm. The mode-dependent loss difference is less than 3 dB for all spatial mode channels over 21 nm spectral bandwidth from 1551 nm to 1572 nm. It’s worth noting that the peak coupling efficiencies of the four spatial channels have a relative wavelength shift of about 45 nm, which is mainly due to the fact that the optimal mode overlap integrals are obtained at different wavelengths for various modes.

%%%%%%%%%%%%%%%%%%%%%%%%%%%%%%%%%%%%%%%%%%%%
% FIGURE 2 - design graph
%%%%%%%%%%%%%%%%%%%%%%%%%%%%%%%%%%%%%%%%%%%%

Because of the arbitrary LP mode evolution from the coherent sum of the vectorial fiber modes and polarization rotation in a circular-core FMF, the field pattern arriving at the FMF end is always uncertain. Therefore, the received field pattern by grating coupler cannot be assumed as a pure LP mode with a perfectly aligned polarization. Nevertheless, our proposed MMGC can convert all the vectorial fiber modes in a two-mode FMF including HE\textsubscript{11-x}, HE\textsubscript{11-y}, TM\textsubscript{01}, TE\textsubscript{01}, HE\textsubscript{21-even}, HE\textsubscript{21-odd} into TE\textsubscript{0} and TE\textsubscript{1} mode on chip. Hence, optical pattern generated by the linear superposition of those vectorial modes at the fiber end can always be coupled and collected using the eight on-chip single-mode waveguides for signal descrambling. Such unique property is confirmed in 3D-FDTD simulation by launching various vectorial mode patterns in a 2-mode FMF and summing the received power on chip (see Supporting Information S3), the obtained coupling efficiencies agree well with the prior chip-to-fiber simulation performance.

\subsection{Mode Independent Mode Size Converter Design}

%%%%%%%%%%%%%%%%%%%%%%%%%%%%%%%%%%%%%%%%%%%%
% FIGURE 3 - Mikaelian lens design
\begin{figure*}[!htp]
  \centering{
  \includegraphics[width = 0.97\linewidth]{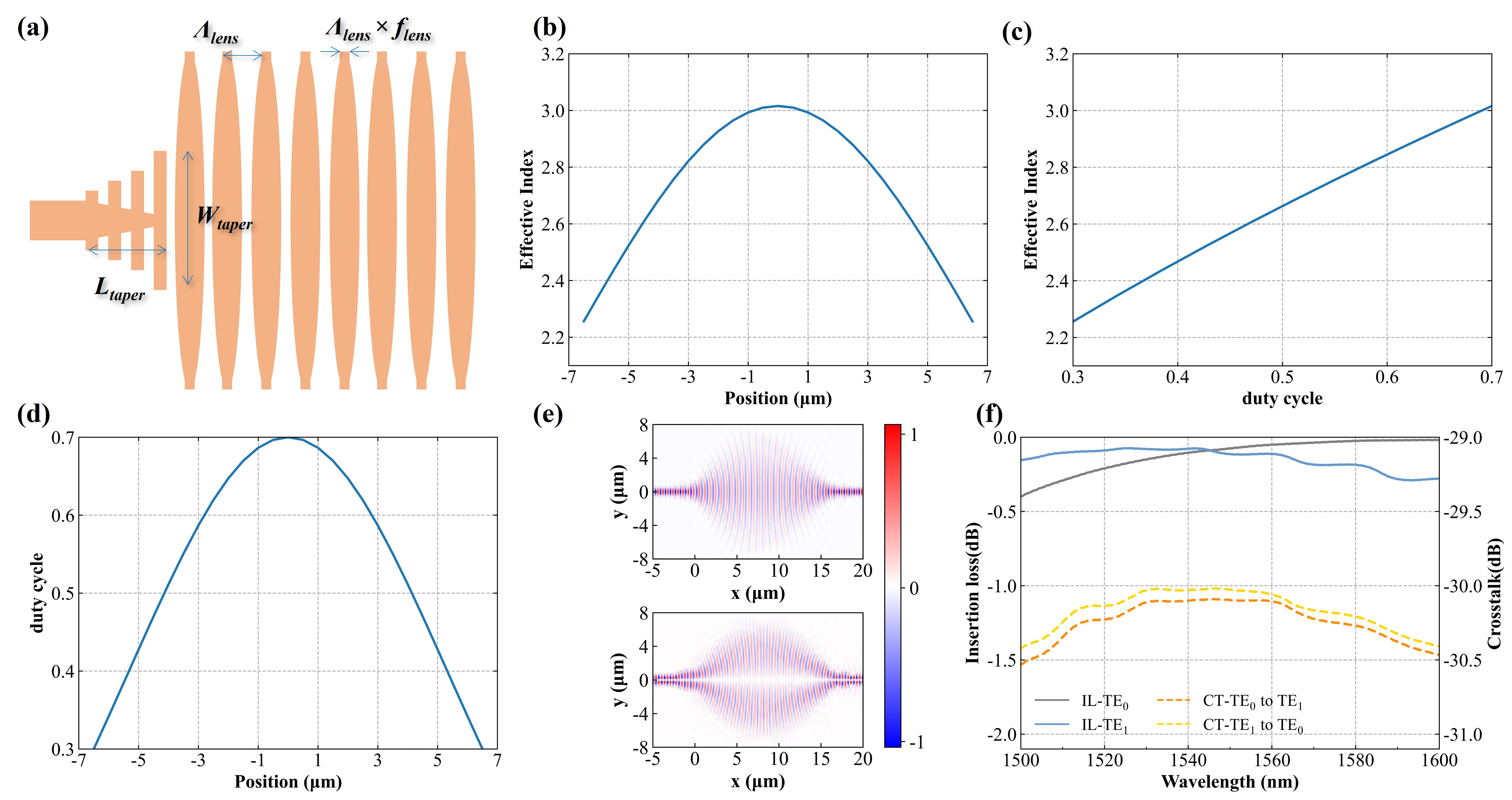}
  } 
    \caption{(a) Schematic of subwavelength Mikaelian lens. (b) Desired refractive index profile matching with a Mikaelian lens. (c) Predicted effective index of the subwavelength grating against duty cycle by the second order EMT. (d) Mapped duty cycle profile of the subwavelength Mikaelian lens. (e) Simulated optical field profiles of the subwavelength Mikaelian lens back-to-back for mode field size conversion. (f) Simulated transmission spectra and inter-modal crosstalk of the subwavelength Mikaelian lens.
  }
 \label{Figure3}
\end{figure*} 
%%%%%%%%%%%%%%%%%%%%%%%%%%%%%%%%%%%%%%%%%%%%

The large refractive index contrast of the SOI platform allows for effective confinement of optical modes within the silicon waveguides, enabling an extremely high level of integration density. Nevertheless, the mode field size also varies considerably between optical fibers and silicon channel waveguides. The highly confined waveguide modes with a typical mode field diameter of $<1\,\mu\text{m}$ is not matched with fiber mode with a typical mode field diameter of $\sim10\,\mu\text{m}$. A long adiabatic linear taper is typically required for mode size conversion. In our case, a linear waveguide taper with a length of 350 $\mu$m is necessary to minimize the transition loss for both the TE\textsubscript{0} and TE\textsubscript{1} modes. The taper length cannot be shortened without violating the adiabatic transition condition. The use of four linearly tapered MSCs for the proposed DE/MUX would cause a large device footprint and waste a lot of valuable chip area.

A compact and mode-independent mode size converter is thus a key step to improve the integration density. Mikaelian lens developed from Maxwell's fish-eye lens using conformal transformation optics, is inherently free of spherical aberrations. Such unique property enables both paraxial and nonparaxial light to be focused at the same point using a single lens, making it perfectly suitable for mode-independent field size conversion. A compact integrated Mikaelian lens on photonic chip can be realized by using the subwavelength grating structures, which have already been widely exploited for many integrated photonic components \cite{cheben_subwavelength_2018,luquegonzalez_ultracompact_2019,liu_arbitrarily_2019,zhang_nonparaxial_2020,xiang_metamaterial-enabled_2022}.

The schematic of subwavelength Mikaelian lens is depicted in Figure \ref{Figure3}a. By changing the duty cycle (defined by the ratio of silicon width to grating period) of the subwavelength silicon gratings, the effective refractive index is manipulated to form a desired refractive index profile of a Mikaelian lens shown in Figure \ref{Figure3}b.

The subwavelength Mikaelian lens is designed to have a width of 13 $\mu$m to match with the mode field diameter of the graded-index FMF. The maximum effective index n\textsubscript{max} is 3.0158 in the center of the subwavelength gratings while the minimum effective index n\textsubscript{min} is 2.2560 at the edge. The subwavelength grating pitch is set as 240 nm while the duty cycle varies continuously from 0.3 to 0.7. To map the effective index profile efficiently, the second order EMT is applied (see Supporting Information S2). Figure \ref{Figure3}c presents the predicted effective index of the subwavelength grating against duty cycle. Effective index profile of the Mikaelian lens can thus be mapped to the duty cycles of the subwavelength gratings as presented in Figure \ref{Figure3}d. A compact subwavelength Mikaelian lens is thus formed for mode-independent size conversion. To mitigate the reflection loss induced by the disparityu in effective indices between the subwavelength lens and the channel waveguide, a subwavelength taper was incorporated at the beginning of the Mikaelian lens in Figure \ref{Figure3}a. The width and length of the subwavelength taper are fine-tuned to 2.5 $\mu$m and 3.6 $\mu$m, respectively to minimize the loss.

Figure \ref{Figure3}e shows the simulated optical field profiles of the designed subwavelength Mikaelian lens back-to-back working for TE\textsubscript{0} and TE\textsubscript{1} modes. A compact and mode-independent field size conversion can be observed. The corresponding transmission spectra and inter-modal crosstalk are summarized in Figure \ref{Figure3}f. A peak conversion efficiency of -0.02 dB and -0.07 dB can be obtained for the TE\textsubscript{0} and TE\textsubscript{1} modes, respectively. Meanwhile, the inter-modal crosstalk is below -30 dB over the whole C band.

%%%%%%%%%%%%%%%%%%%%%%%%%%%%%%%%%%%%%%%%%%%%
% FIGURE 4 - SEM
\begin{figure*}[!htp]
  \centering{
  \includegraphics[width = 0.97\linewidth]{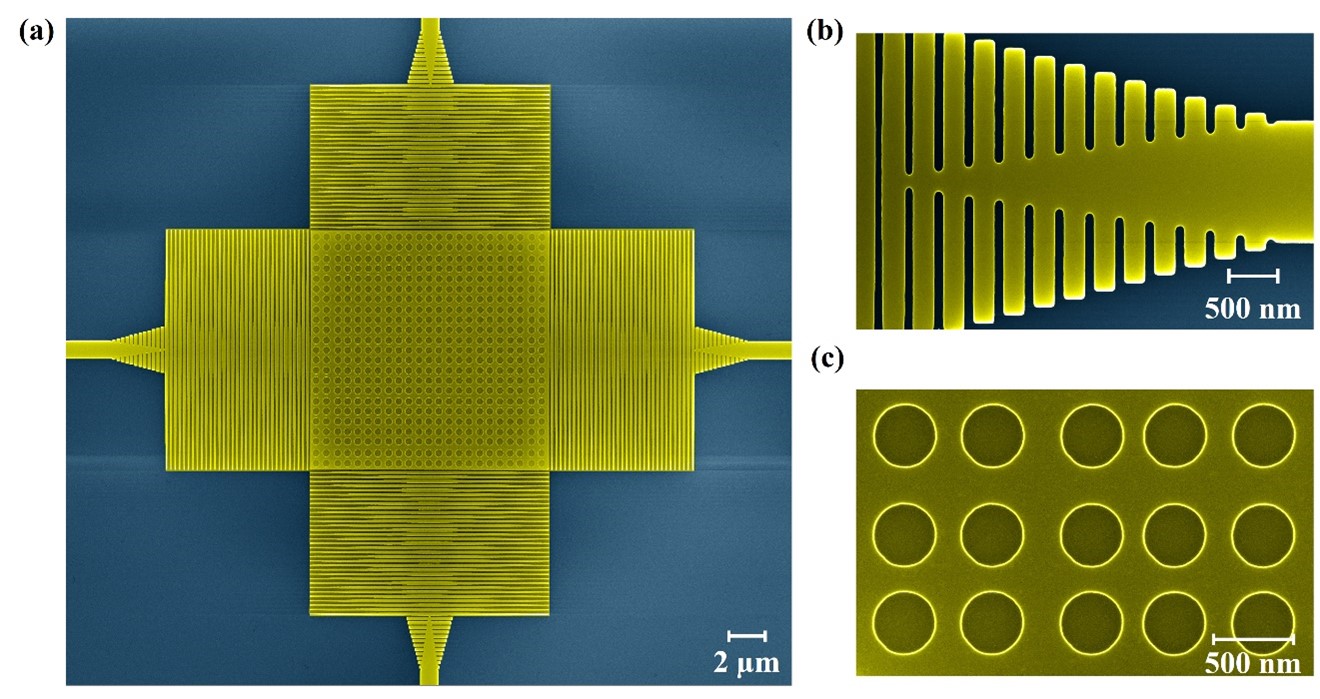}
  } 
    \caption{(a) Scanning electron microscope (SEM) images of the fabricated MMGC and MSCs. (b) Zoom-in view of the fabricated subwavelength taper and Mikaelian lens. (c) Zoom-in view of the 70-nm shallow etched holes in the MMGC.
  }
 \label{Figure4}
\end{figure*} 
%%%%%%%%%%%%%%%%%%%%%%%%%%%%%%%%%%%%%%%%%%%%

\section*{Fabrication and Experimental Results}
The proposed integrated multimode DE/MUX is fabricated on SOI wafer with a 220 nm thick top silicon layer. The buried-oxide layer has a thickness of 2 $\mu$m. Electro-beam lithography is used to define the device patterns, followed by dry reactive-ion etching process with a shallow etch depth of 70 nm and a full etch. Silicon dioxide (SiO\textsubscript{2}) with a thickness of 1.2 $\mu$m is used as the top cladding to protect the passive photonic circuits. Metallization is performed using high-resistance titanium-tungsten alloy (TiW) for local heat generation and aluminum for electrical signal routing. A 300-nm thick SiO\textsubscript{2} passivation layer is used and selectively etched later to create windows over the aluminum pads for probing.

Scanning electron microscope (SEM) images of the fabricated MMGC and MSCs are presented in Figure \ref{Figure4}a-c. The total footprint in Figure \ref{Figure4}a is only 35×35 $\mu$m\textsuperscript{2}. Compared with the case by using the linear adiabatic tapers (713×713 $\mu$m\textsuperscript{2}) shown in the Figure S4a, the total footprint can be reduced by more than 400-fold.

%%%%%%%%%%%%%%%%%%%%%%%%%%%%%%%%%%%%%%%%%%%%
% FIGURE 5 - EXP results
\begin{figure*}[!htp]
  \centering{
  \includegraphics[width = 0.97\linewidth]{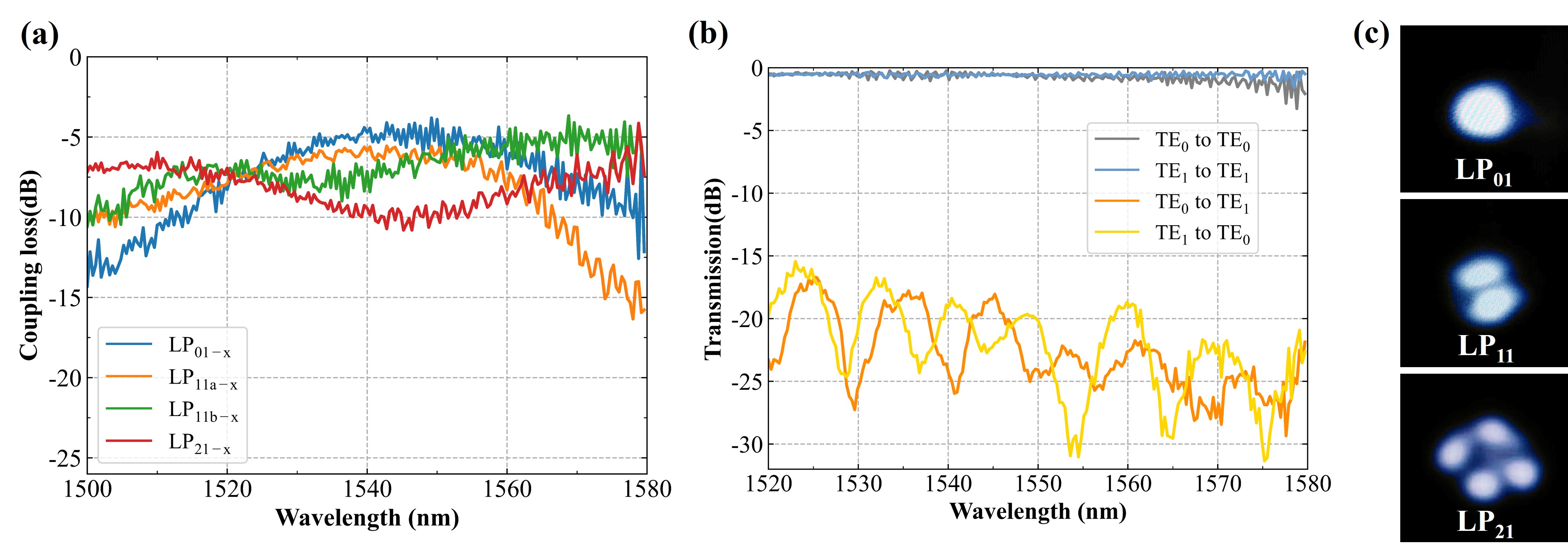}
  } 
    \caption{(a) Normalized coupling loss spectra of the MMGC in the experiment for mode LP\textsubscript{01-x}, LP\textsubscript{11a-x}, LP\textsubscript{11b-x}, LP\textsubscript{21b-x} for a four-mode FMF. (b)Measured insertion loss and crosstalk spectra for the subwavelength Mikaelian lens. (c) Mode field profiles of the FMF captured by an infrared camera when different modes are selectively launched.
  }
 \label{Figure5}
\end{figure*} 
%%%%%%%%%%%%%%%%%%%%%%%%%%%%%%%%%%%%%%%%%%%%

\begin{table*}
\caption{Integrated DE/MUX by Diffraction Gratings for MMFs}
\label{table 1}
\resizebox{1.0\linewidth}{!}{
\begin{tabular}{ccccccccccccc}
\hline
Ref. & Multimode fiber type & Num. of spatial channels & Experimental minimum coupling loss [dB] & Mode dependent loss [dB] & Footprint [MMGs and MSCs] $\mu$m\textsuperscript{2}\\
\hline
\cite{yang_multi-dimensional_2022}  & Rectangular core & 4  & TE\textsubscript{00}-TE\textsubscript{30}: <-5.5 dB & < 2.5 dB  & -\\
\cite{koonen_silicon_2012}  & Circular core & 6 & LP\textsubscript{01}-LP\textsubscript{11} <-20 dB & ~ 5 dB & -\\
\cite{ding_silicon_2013}  & Circular core & 6   & - & < 2 dB  & -\\
\cite{tong_efficient_2019} & Circular core  & 4  & LP\textsubscript{01}: −4.9 dB and LP\textsubscript{11a}: −6.1 dB
  & $\sim1.2\,\text{dB}$ & 630×630\\
\cite{watanabe_coherent_2020} & Circular core  & 6  & LP\textsubscript{01}: −5.2 dB,
LP\textsubscript{11}: −9.0 dB and
  & $\sim4.2\,\text{dB}$ & >200×200\\
This work & Circular core  & 8  & LP\textsubscript{01}: −3.8 dB,
LP\textsubscript{11a}: −5.5 dB,	LP\textsubscript{11b}: −3.6 dB and LP\textsubscript{21b}: −4.1 dB
  & Wavelength dependent  & 35×35\\
\hline
\end{tabular}
}
\end{table*}

The coupling loss of the fabricated MMGC was characterized by measuring the fiber-chip-fiber transmission using a tunable continuous-wave laser source and a power meter. A SMF with single-mode grating coupler is used at the input side. The polarization of the input mode was aligned by a three-paddle mechanical polarization controller so that all the transmission for x-polarization is maximized. An integrated Y-branch is used as the 50:50 power splitter, followed by 150-µm long waveguide heaters for phase tuning. The optical signals are sent to the ADCs, MSCs and MMGC with a 4-mode FMF at the output to collect optical power. The heater injection current of the TE\textsubscript{0} mode can be tuned to maximize the transmission at 1550 nm or 1570 nm, to selectively launch LP\textsubscript{01} or LP\textsubscript{11b}, respectively. Similarly, LP\textsubscript{11} and LP\textsubscript{21b} can be selectively excited using the TE\textsubscript{1} mode by the same approach. Figure \ref{Figure5}a shows the normalized coupling loss spectra of the MMGC using a four-mode FMF. LP\textsubscript{01} and LP\textsubscript{11a} have a peak coupling efficiency of -3.8 dB and -5.5 dB at 1549 nm and 1542 nm, respectively. LP\textsubscript{11b} and LP\textsubscript{21b} have a peak coupling efficiency of -3.6 dB and -4.1 dB at 1578nm and 1569nm respectively. The observed 1-dB spectral bandwidth is about 20 nm for all the spatial channels, which is narrower than the simulation results. It is mainly due to that the relative phase shift is only optimized for the center wavelength when performing the wavelength scan. Hybrid modes are thus excited at the edges of the wavelength scanning range. The presence of mode dependent loss can lead to a decrease of the transmission power and the spectral bandwidth, correspondingly.

The insertion loss and crosstalk of the integrated subwavelength Mikaelian lens are also characterized by measuring the fiber-chip-fiber transmission. Due to the low insertion loss of the proposed MSC, four Mikaelian lens connected back-to-back (see Figure S4b) are used to obtain the normalized insertion loss and crosstalk shown in Figure \ref{Figure5}b. The measured minimum insertion loss of the TE\textsubscript{0} and TE\textsubscript{1} modes are -0.24 dB at 1540nm and -0.25 dB at 1573 nm, respectively. Inter-modal crosstalk levels can be well suppressed to less than -16 dB. The increased spectral ripples around 1570 nm are mainly due to limited sensitivity of our power meter when using the lossy single mode grating couplers, which are originally designed to be centered at 1545 nm.

To confirm the selective excitation of 8 spatial channels in FMF as illustrated in Figure \ref{Figure1}c, we use an infrared camera with a 10× microscope objective to record the output field profile of the FMF. Figure \ref{Figure5}c present the intensity profiles of the mode group LP\textsubscript{01}, LP\textsubscript{11}, and LP\textsubscript{21}, which confirms the selective launching ability of our design. Since the LP mode evolution and polarization rotation in a circular-core FMF are unpredictable, the FMF uses a three-paddle mechanical polarization controller to capture pure LP\textsubscript{11} and LP\textsubscript{21} at the fiber end as much as possible.

Table \ref{table 1} summarizes integrated mode DE/MUX by the grating coupling approach for multimode fibers in the past decade. Compared with prior arts, our work for the first time shows an ultra-compact, efficient, and multichannel solution on chip to selectively launch 8 spatial channels in the FMF. 

\section*{Conclusion}
To summarize, an ultra-compact and efficient integrated multichannel mode multiplexer in silicon for few-mode fibers is demonstrated. Selectively launching of 8 spatial channels with a minimum coupling loss of -3.8 dB, -5.5 dB, -3.6 dB, -4.1 dB for LP\textsubscript{01}, LP\textsubscript{11a}, LP\textsubscript{11b}, and LP\textsubscript{21b} modes are demonstrated in experiment. Since a low coupling loss can be maintained for all the degenerate modes regardless of the mode or polarization change during transmission in a two-mode FMF, the proposed design can also work in a DEMUX where signal descrambling is required to eliminate modal crosstalk. More than that, the demonstrated design is very compact by using the subwavelength Mikaelian lens for mode independent field size conversion. The total footprint of the MMGC and MSCs can be shrunk by more than 400-fold to only 35×35 $\mu$m\textsuperscript{2}, paving the way for the densely integrated mode multiplexed systems.

\vspace{0.2cm}

\noindent \textbf{Supporting Information}:
Supporting Information is available from the Wiley Online Library or from the author.

\vspace{0.2cm}

\noindent \textbf{Acknowledgements}:
This work was funded by the Guangzhou-HKUST(GZ) Joint Funding Scheme 2023A03J0159, Start-up fund from the Hong Kong University of Science and Technology (Guangzhou) and Hong Kong RGC GRF grant 14203620. The authors acknowledge the Novel IC Exploration (NICE) Facility of HKUST(GZ) for device measurement and Applied Nanotools Inc. for device fabrication. Dr. Yi Wang and Mr. Yue Qin are acknowledged for technical assistance.

\bibliographystyle{naturemag}
\bibliography{Reference}

\end{document}

\title{Supporting Information for:\\Ultra-compact and efficient integrated multichannel mode multiplexer in silicon for few-mode fibers}

\author{Wu Zhou$^{1}$, 
        Zunyue Zhang$^{2,3}$, 
        Hao Chen$^{1}$, 
        Hon Ki Tsang$^{2}$, 
        and Yeyu Tong$^{1,\dagger}$}
\affiliation{
$^1$Microelectronic Thrust, Function Hub, The Hong Kong University of Science and Technology (Guangzhou), Guangdong, PR China\\
$^2$Department of Electronic Engineering, The Chinese University of Hong Kong, Shatin, New Territories, Hong Kong, PR China\\
$^3$School of Precision Instrument and Opto-Electronics Engineering, Tianjin University, Tianjin, PR China\\
}
\maketitle

\noindent{\textbf{\large{Contents}}}

\noindent{\textbf{Supplementary Notes:}}\\
\noindent{1. Asymmetrical Directional Couplers (ADCs) Design}\\
2. Effective Medium Theory (EMT)\\
3. Multimode Grating Coupler (MMGC) Interfaced with a 2-mode FMF

%%%%%%%%%%%%%%%%%%%%%%%%%%%%%%%%%%%%%%%%%%%%
% FIGURE S1 - mode field profiles
\renewcommand{\thefigure}{S\arabic{figure}}
\begin{figure*}[!htp]
  \centering{
  \includegraphics[width = 0.97\linewidth]{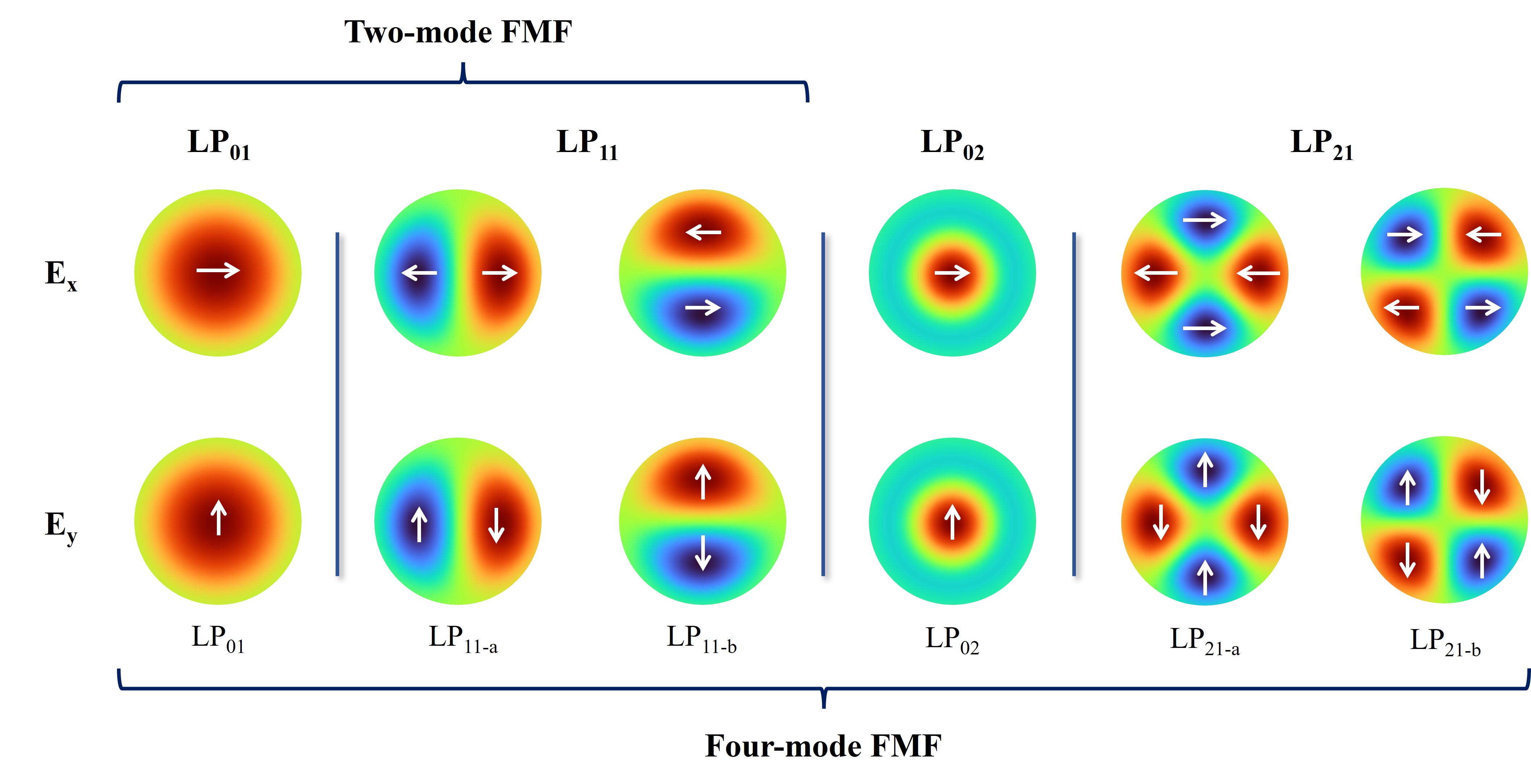}
  } 
    \caption{Mode field profiles of the supported LP modes in a two-mode FMF and a four-mode FMF.
  }
 \label{FigureS1}
\end{figure*} 
%%%%%%%%%%%%%%%%%%%%%%%%%%%%%%%%%%%%%%%%%%%%

%%%%%%%%%%%%%%%%%%%%%
\section*{\textbf{Supplementary Note 1. Asymmetrical Directional Couplers (ADCs) Design}}
The on-chip two-mode multiplexer demonstrated in this work is based on tapered asymmetrical directional couplers (ADCs). As shown in Figure \ref{FigureS2}a, a tapered directional coupler is used in the coupling region in order to relax the fabrication tolerance of the normal ADCs. The ADC is optimized to maximize the transmission efficiency, The waveguide width w\textsubscript{1}, w\textsubscript{2a}, and w\textsubscript{2b} used are 0.45 \(\mu\)m, 0.902 \(\mu\)m, and 0.962 \(\mu\)m, respectively. The waveguide gap g is set as 0.2 \(\mu\)m and coupler length L is 33.6 \(\mu\)m. The simulated coupling efficiency and modal crosstalk are shown in Figure \ref{FigureS2}b. Figure \ref{FigureS2}c is a zoom-in scanning electron microscope image of the fabricated ADC.

%%%%%%%%%%%%%%%%%%%%%%%%%%%%%%%%%%%%%%%%%%%%
% FIGURE S2 - ADC
\renewcommand{\thefigure}{S\arabic{figure}}
\begin{figure*}[!htp]
  \centering{
  \includegraphics[width = 0.97\linewidth]{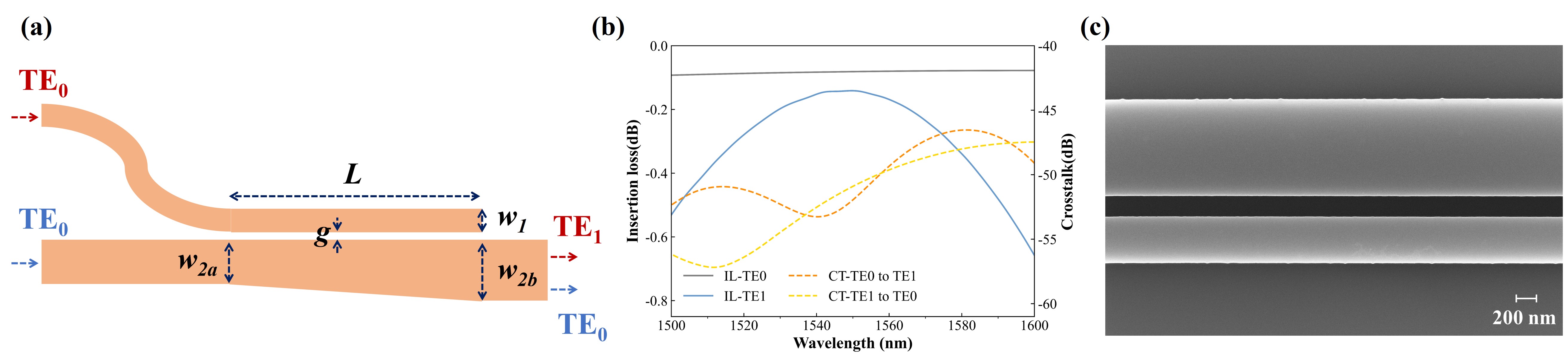}
  } 
    \caption{(a) Schematic diagram of 2-mode tapered ADC for on-chip mode de/multiplexing of the TE\textsubscript{0} and TE\textsubscript{1} mode. (b) microscopic image the back-to-back ADCs for loss and modal crosstalk characterization. (c) Scanning microscopic images of the fabricated directional coupler with a waveguide gap of 200 nm.
  }
 \label{FigureS2}
\end{figure*} 
%%%%%%%%%%%%%%%%%%%%%%%%%%%%%%%%%%%%%%%%%%%%

\section*{\textbf{Supplementary Note 2. Effective Medium Theory (EMT)}}
Subwavelength periodic structures refer to patterns that have a small enough spacing. The effective refractive index of a material can be manipulated by changing the duty cycle or filling factor of the subwavelength grating, which offers a practical method to create a desired refractive index. The refractive index of subwavelength structures can be calculated by using the second order EMT, as denoted by Equations \eqref{eq:S1}-\eqref{eq:S2}. \( n_{\text{TM}}^{(0)} \) and \( n_{\text{TE}}^{(0)} \) are the estimated effective indices of the transverse-magnetic (TM) mode and TE mode using the zeroth-order approximation given by Equations \eqref{eq:S3}-\eqref{eq:S4}. The refractive index of silicon and cladding silicon dioxide are \( n_{\text{silicon}} \) and \( n_{\text{oxide}} \), respectively. \(\lambda\) is the center optical wavelength, \(\Lambda_{\text{avg}}\) and \(f_{\text{avg}}\) are the grating period and filling factor (defined by the ratio of silicon dioxide width to grating period).

\renewcommand{\theequation}{S\arabic{equation}}
\begin{equation}
n_{\text{TM}}^{(2)} = n_{\text{TM}}^{(0)} \left[1 + \frac{\pi^2}{3} \left(\frac{\Lambda_{\text{avg}}}{\lambda}\right)^2 f_{\text{avg}}^2 (1-f_{\text{avg}})^2 \times \frac{(n_{\text{silicon}}^2 - n_{\text{oxide}}^2)^2}{(n_{\text{TM}}^{(0)})^2} \right]^{1/2}
\label{eq:S1}
\end{equation}

\begin{equation}
n_{\text{TE}}^{(2)} = n_{\text{TE}}^{(0)} \left[1 + \frac{\pi^2}{3} \left(\frac{\Lambda_{\text{avg}}}{\lambda}\right)^2 f_{\text{avg}}^2 (1-f_{\text{avg}})^2 \times \frac{(n_{\text{silicon}}^2 - n_{\text{oxide}}^2)^2 (n_{\text{TM}}^{(0)})^2 \left(\frac{(n_{\text{TE}}^{(0)})^2}{n_{\text{silicon}}^2 n_{\text{oxide}}^2}\right)^2}{(n_{\text{TE}}^{(0)})^2} \right]^{1/2}
\label{eq:S2}
\end{equation}

\begin{equation}
n_{\text{TM}}^{(0)} = \left[f_{\text{avg}} n_{\text{oxide}}^2 + (1 - f_{\text{avg}}) n_{\text{silicon}}^2\right]^{1/2}
\label{eq:S3}
\end{equation}

\begin{equation}
\frac{1}{n_{\text{TE}}^{(0)}} = \left[\frac{f_{\text{avg}}}{n_{\text{oxide}}^2} + \frac{(1 - f_{\text{avg}})}{n_{\text{silicon}}^2}\right]^{-1/2}
\label{eq:S4}
\end{equation}

%%%%%%%%%%%%%%%%%%%%%%%%%%%%%%%%%%%%%%%%%%%%
% FIGURE S3 - Coupling loss
\renewcommand{\thefigure}{S\arabic{figure}}
\begin{figure*}[!htp]
  \centering{
  \includegraphics[width = 0.97\linewidth]{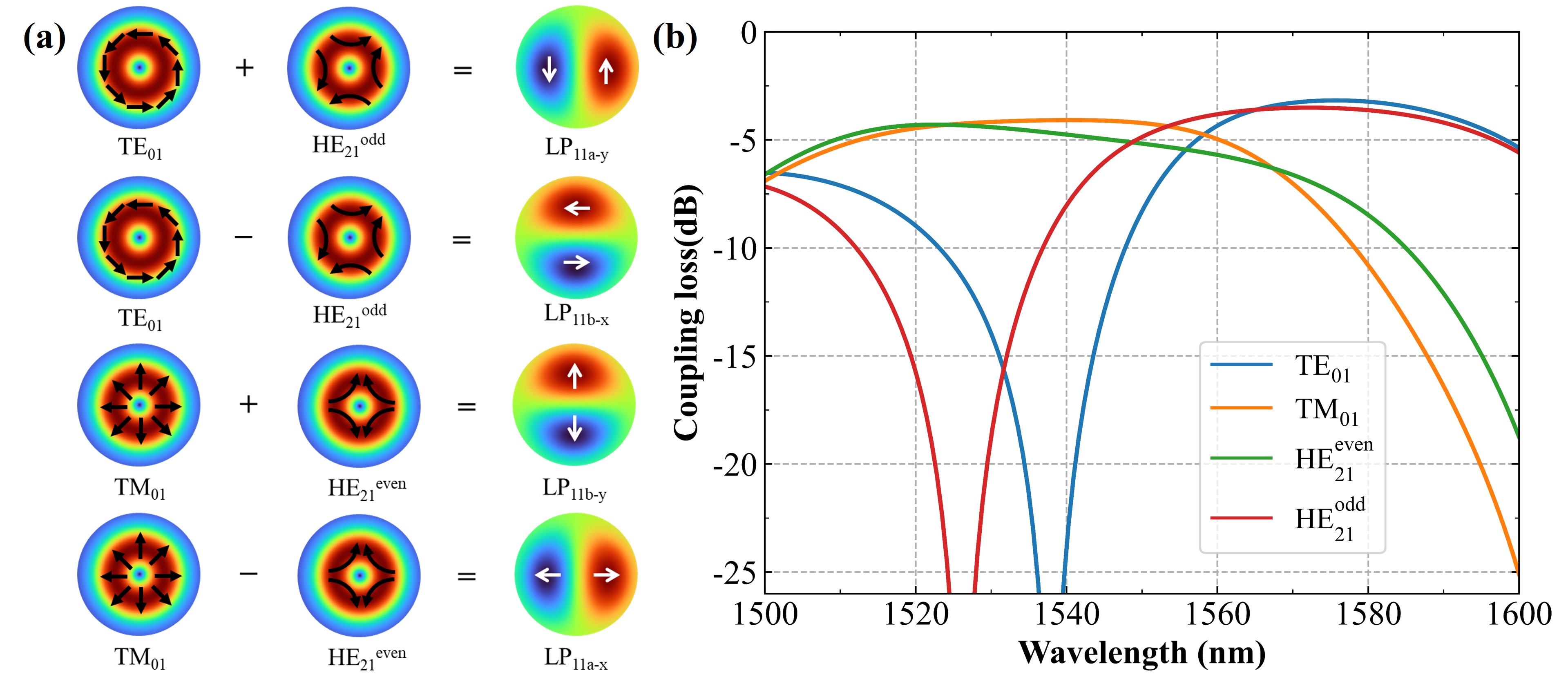}
  } 
    \caption{(a) Illustration of LP11 modes formed by the linear superpositions of vectorial fiber modes in a two-mode FMF. (b) Simulation results of the coupling loss by launching various vectorial fiber mode including TM\textsubscript{01}, TE\textsubscript{01}, HE\textsubscript{21-even}, and HE\textsubscript{21-odd} and the total received power into TE\textsubscript{0} and TE\textsubscript{1} modes on chip.
  }
 \label{FigureS3}
\end{figure*} 
%%%%%%%%%%%%%%%%%%%%%%%%%%%%%%%%%%%%%%%%%%%%

\section*{\textbf{Supplementary Note 3. Multimode Grating Coupler (MMGC) Interfaced with a 2-mode FMF)}}
The intensity distribution at the FMF end is the coherent sum of vectorial modes, which is uncertain in practice. The vectorial modes in a two-mode FMF includes HE\textsubscript{11-x} (LP\textsubscript{01-x}), HE\textsubscript{11-y} (LP\textsubscript{01-y}), TM\textsubscript{01}, TE\textsubscript{01}, HE\textsubscript{21-even}, and HE\textsubscript{21-odd}. The fundamental modes HE\textsubscript{11-x} and HE\textsubscript{11-y} are not considered here which are the same the 2D grating coupler for a single mode fiber. Each of the pure scalar LP11 modes is the linear superpositions of appropriate pairs of the vectorial fiber modes as depicted in Figure \ref{FigureS3}a. Our proposed design shown in Figure 1a can work for all the degenerate modes in a two-mode FMF and convert them into eight fundamental TE modes on the photonic chip. Figure \ref{FigureS3}b presents the fiber-to-chip coupling loss spectra by summing the received power into TE\textsubscript{0} and TE\textsubscript{1} modes, when different vectorial modes are selectively launched in a 2-mode FMF. TE\textsubscript{01} and HE\textsubscript{21-odd} have a peak coupling efficiency of -3.17 dB at 1575 nm, -3.51 dB at 1572 nm respectively. The linear superposition of TE\textsubscript{01} and HE\textsubscript{21-odd} results in LP\textsubscript{11b-x} whose coupling efficiency is shown in Figure 2c. The TM\textsubscript{01} and HE\textsubscript{21-even} have a peak coupling efficiency of -4.08 dB at 1540 nm, -4.29 dB at 1522 nm respectively. The linear superposition of TM\textsubscript{01} and HE\textsubscript{21-even} forms the LP\textsubscript{11a-x} with coupling efficiency presented in Figure 2c. The simulated efficiencies agree well between the scenarios of chip-to-fiber coupling and fiber-to-chip coupling shown in Figure 2c and Figure \ref{FigureS3}b. The small discrepancy in peak efficiency may be caused by the simulated mesh size difference and overlap integration error.

%%%%%%%%%%%%%%%%%%%%%%%%%%%%%%%%%%%%%%%%%%%%
% FIGURE S4 - Microscopic image of the MMGC
\renewcommand{\thefigure}{S\arabic{figure}}
\begin{figure*}[!htp]
  \centering{
  \includegraphics[width = 0.97\linewidth]{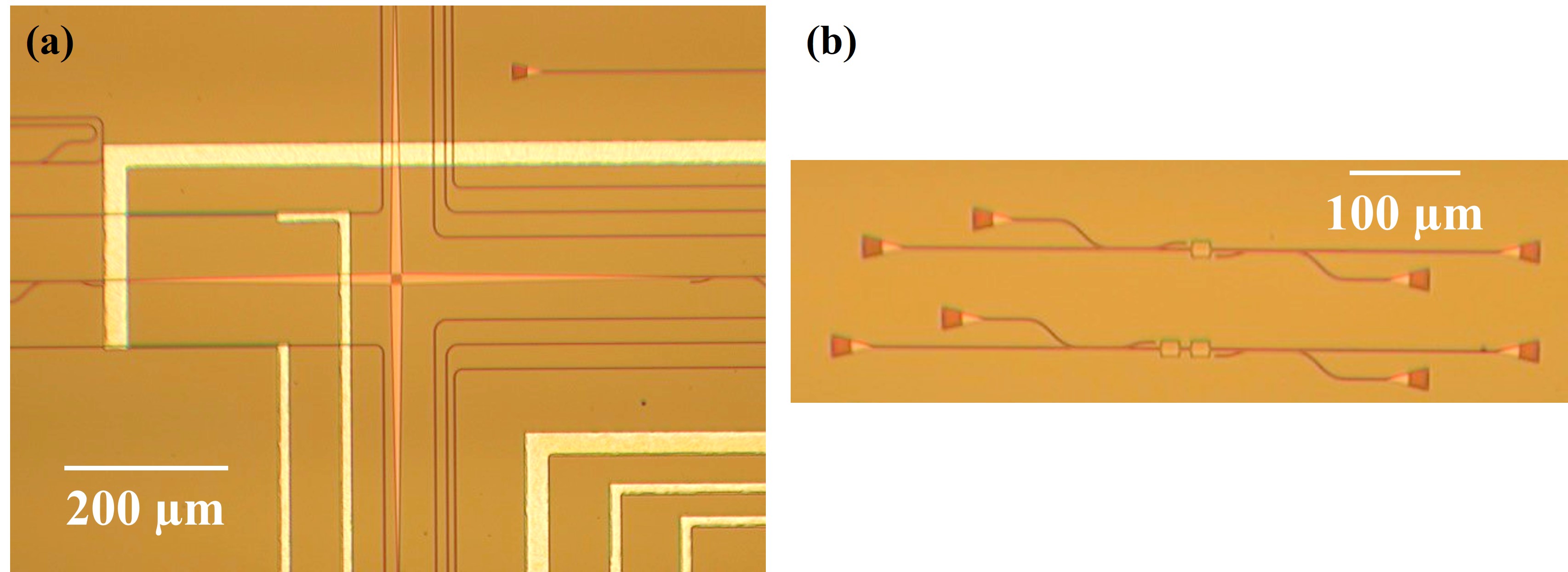}
  } 
    \caption{(a) Microscopic image of the MMGC and four linearly adiabatic tapers. (b) Microscopic image of two and four Mikaelian lens connected back-to-back.
  }
 \label{FigureS4}
\end{figure*} 
%%%%%%%%%%%%%%%%%%%%%%%%%%%%%%%%%%%%%%%%%%%%